# Impact of silver addition on room temperature magneto-resistance in $La_{0.7}Ba_{0.3}MnO_3$ (LBMO): $Ag_x$ (x = 0.0, 0.1, 0.2, 0.3, 0.4)

Rahul Tripathi, V.P.S. Awana*, and H. Kishan
National physical Laboratory, Dr. K.S. Krishnan Marg, New Delhi-110012, India

S. Balamurugan
Advanced nano-materials Laboratory, National Institute for Material Science (NIMS), Tsukuba, Ibaraki 305-0044, Japan

G.L. Bhalla
Department of Physics and Astrophysics, Delhi University, New Delhi, India

$La_{0.7}Ba_{0.3}MnO_3$ (LBMO):$Ag_x$ (x = 0.0, 0.1, 0.2, 0.3, and 0.4) composites are synthesized by solid-state reaction route, the final sintering temperatures are varied from 1300 (LBMO1300Ag) to 1400 $^0$C (LBMO1400Ag), and their physical properties are compared as a function of temperature and Ag content. All samples are crystallized in single phase accompanied by some distortion in main structural phase peaks at higher angles with increase in silver content. Though the lattice parameters (a, c) decrease, the b increases slightly with an increase in Ag content. The scanning electron micrographs (SEM) showed better grains morphology in terms of size and diffusion of grain boundaries with an increase in Ag content. In both LBMO1300Ag and LBMO1400Ag series the metal insulator transition ($T^{MI}$) and accompanied paramagnetic-ferromagnetic transition ($T_C$) temperatures are decreased with increase in Ag content. The sharpness of MI transition, defined by temperature coefficient of resistance (TCR), is improved for Ag added samples. At a particular content of Ag(0.3), the $T^{MI}$ and $T_C$ are tuned to 300K and maximum magneto-resistance at 7Tesla applied field ($MR^{7T}$) of up to 55% is achieved at this temperature, which is more than double to that as observed for pure samples of the both 1300 and 1400 $^0$C series at same temperature. The $MR^{7T}$ is further increased to above 60% for LBMOAg(0.4) samples, but is at 270K. The $MR^{7T}$ is measured at varying temperatures of 5, 100, 200, 300, and 400K in varying fields from ±7 Tesla, which exhibits U and V type shapes. Summarily, the addition of Ag in LBMO improves significantly the morphology of the grains and results in better physical properties of the parent manganite system.

*: Corresponding Author
Dr. V.P.S. Awana
Room 109,
National Physical Laboratory, Dr. K.S. Krishnan marg, New Delhi-110012, India
Fax No. 0091-11-25626938: Phone no. 0091-11-25748709
e-mail-awana@mail.nplindia.ernet.in: www.freewebs.com/vpsawana/

## INTRODUCTION

Manganites with general formula $A_{1-x}A'_xMnO_3$ ($ABO_3$ type), where A is the rare earth element and A′ is the divalent cation with $0.0 \le x \le 1.0$, have already been studied extensively for their technically important properties viz. colossal magneto-resistance (CMR) & Magneto caloric effect along with their rich fundamental physics [1-6]. This class of compounds exhibits a broad spectrum of magnetic phases like anti-ferromagnetic insulator, ferromagnetic metallic and paramagnetic insulator states [5]. Different magnetic and electrical phases in these compounds have their origin in charge(s), spin and orbital interplay which can be tuned by divalent as well as monovalent substitution at rare earth site [3-5]. Variation in ionic radii of A site (rare earth/divalent cation) brings substantial changes in the structural parameters of these compounds [4]. Transport properties in these compounds are governed through intra-grain and inter-grain channels, thus grain growth and connectivity plays a decisive role. In our previous work on $La_{0.7}Ca_{0.3}MnO_3$ (LCMO):$Ag_x$/Indium composites[7-8], we found that the addition of silver in LCMO improved the grain growth as well as their connectivity, which resulted in high value (>15%) of temperature coefficient of resistance (TCR) and high MR [7]. High value of TCR has its applications in infrared detectors [6]. Further for practical applications of manganites the high TCR and MR need to be achieved at/or above room temperature. In case of LCMO:$Ag_x$ composites even though high TCR of above 15% was achieved but at a temperature (280 K) lower than the room temperature [7]. Since for $La_{0.7}Ba_{0.3}MnO_3$(LBMO), the transition temperature ($T_C$) [where the maximum TCR and MR can be achieved] is above room temperature [3,4], we have synthesized its silver added composites for better room temperature performance. Here in this article, we report the effect of different sintering temperatures on $La_{0.7}Ba_{0.3}MnO_3$(LBMO):$Ag_x$ (x = 0.1, 0.2, 0.3, 0.4). The physical properties of the system in terms of room temperature MR and TCR along with its ferromagnetic order are reported and discussed. With progressive addition of Ag in LBMO, the high MR of up to 60% is achieved at room temperature with improved TCR. Ag addition is found to improve the grain morphology of the LBMO dramatically and hence better physical properties.

## EXPERIMENTAL

$La_{0.7}Ba_{0.3}MnO_3$ (LBMO):$Ag_x$(x = 0.0, 0.1, 0.2, 0.3, 0.4) composites were synthesized by solid-state reaction route using high purity powders of $La_2O_3$, $BaCO_3$, $MnO_2$ and Ag in stoichiometric ratio and ground thoroughly. All the mixed powders were calcined four times between 900 ºC till 1200 $^oC$ with the step of $100^oC$, for 24 hours, which was followed by natural cooling to room temperature. These calcined powders were palletized using hydraulic press and annealed in air for 48 hours in two batches one at 1300 ºC (LBMO1300Ag) and the other at 1400 ºC (LBMO1400Ag) respectively .In the end these final pellets were annealed in the flow of oxygen at 800 ºC for 48 hours and subsequently slow cooled to room temperature. The structure and phase purity of the $La_{0.7}Ba_{0.3}MnO_3$ (LBMO):$Ag_x$ composites were checked by powder

X–ray diffraction (XRD) using Ni–filtered Cu $K_{\alpha}$ radiation. Magnetization measurements were carried between 5 K and 400 K under the applied field of 1 Tesla. Isothermal magnetization curves were obtained with applied fields up to ±5 Tesla at fixed (different) temperatures (T = 5, 100, 200, 300 and 400 K). The transport and magneto-transport measurements were carried out in a commercial apparatus (PPMS-Quantum Design) between 5 K and 400 K in magnetic fields of up to 7 Tesla. SEM studies were carried out on these samples using a Leo 440 (Oxford Microscopy: UK) instrument.

## RESULTS & DISCUSSIONS

The X-ray diffraction pattern of $LBMOAg_x$ (x = 0.0, 0.1, 0.2, 0.3, 0.4) sintered at 1300ºC (LBMO1300Ag) & 1400ºC (LBMO1400Ag) are shown in Figure 1(a), 1(b). These figures show that all our samples are crystallized into single-phase [3,4, 9,10], except few un-reacted BaO lines (marked as * in pattern) for some of LBMO1300Ag samples. Also there is no phase change in LBMO (pure) sintered at $1300^{o}$ C or $1400^{o}$ C, that means that the structural properties remains the same even if we synthesized it at higher sintering temperature, please see inset Fig. (1b). The higher angle peaks between 66 and 69 does not exhibit any sign of orthorhombic (*pbnm*) to rhombohedral (*R*3*C*) phase transformation [11]. Hence the crystalline structure considered is orthorhombic in both LBMOAg-1300/1400 $^{0}$C samples, with *pbnm* space group. The indexing of various {hkl} planes are shown on the respective patterns. Note that only the main plane {hkl}are given and due to lack of space the corresponding splitted (*pbnm* space group) peaks are not marked, though they were seen clearly in zoomed patterns as {020}/{200}, {202}/{022}, {312}/{123} etc. On the other hand with successive addition of Ag the space group is seemingly changing, please see inset Fig. 1(a) [12]. The role of Ag addition in manganites in terms of the complicated structural phase transformations is yet warranted. The calculated lattice parameters based on *pbnm* space group are plotted in figure 1(c). The lattice parameters are a = 5.52(2), b = 5.53(8) and c = 7.86(7) for LBMO1400 pure, which get monotonically decreased to a = 5.51(4) and c= 7.80(6) but increased to b = 5.53(6) for LBMO1400Ag0.4 sample. The variation of lattice parameters for LBMO1300Ag is also seen to be similar to that as for LBMO1400Ag. The variation of lattice parameters for LBMO1400Ag series is plotted in Fig. 1(c). Decrease in a and c-parameters indicate towards the lower ionic size Ag substitution at relatively bigger La or Ba site in the doped system. On the other hand increase in b-parameter is indicative of a possible change in oxygen content with Ag substitution [3,4], this has earlier been observed in case of YBCO:Ag high $T_c$ superconducting systems as well [13]. The volume of the unit-cell decreases with increase in Ag content, indicating the substitution of smaller size Ag at bigger La or Ba in LBMO.

Figure 1(d) & 1(e) shows the SEM (scanning electron microscope) pictures of LBMO1400pure & LBMO1400Ag0.4 samples with same magnification. It is seen that the grain size has improved dramatically for Ag0.4 in comparison to that of the pristine sample. Further the grain boundaries look like as diffused into the main grains. This could dramatically improve the physical properties of the doped system in terms of its improved screening current effective surface area and grains connectivity. EDX analyses of LBMO1300Ag & LBMO1400Ag series showed only a small quantity of Ag present in doped the system. So one can conclude that in the case

of LBMOAg addition of silver has enhanced the grain size and the grains connectivity.

The variation of normalized resistivity ($\rho_{(T)}/\rho_{400K}$) with temperature for LBMO1300Ag & LBMO1400Ag composites are shown in figures 2(a) & 2(b). These figures depict that pristine sample (LBMO1300 & LBMO1400) are insulating above their metal-insulator transition temperature ($T^{MI}$) of 330 K and 335 K respectively. Below transition temperature both samples are showing usual metallic behavior with an upturn below 30 K. $\rho$(T) plots of pure samples(x=0 ) are exhibiting two peak character. With silver doping (x = 0.1, 0.2, 0.3, 0.4) in LBMO, instead of double peak only single peak is observed in $\rho$(T) curves i.e. secondary peak vanishes also $T^{MI}$ shifts to lower temperatures. For LBMO1300pure the $T^{MI}$ is 330 K, which shows a significant shift (decrement) of 65 K for LBMO1300$Ag_{0.4}$. Generally shift in the transition temperature is explained in two possible ways: Either there has been an oxygen deficiency in the compound or substitution may have occurred into the lattice [3,4, 11,12, 14]. In our case, samples are showing lower temperature shift in $T^{MI}$. Here the possibility of oxygen deficiency with silver content is low since manganites reduces its oxygen content only under rigorous conditions (of temperature and pressure) and all samples in the present case are synthesized/oxygeneted under the same condition. One can also inquire into the chances of formation of silver oxide where silver can act as oxygen trapper i.e. silver trap oxygen from the lattice. But since foreign grains of silver based compounds are not seen in SEM or XRD, hence this factor is excluded as a cause for the shifting of $T^{MI}$. So the reason for the shift in $T^{MI}$ for system under study may have its origin in substitution of silver at La/Ba site and hence the change in oxygen content. This is consistent with the decreasing a and c lattice parameters with Ag addition in LBMO, see Fig. 1(c). Overall we can say that in Ag composites on one hand silver is decreasing $T^{MI}$ but on the other hand improving the grain size/connectivity. Samples sintered at higher temperature (LBMO1400Ag) follows the same trend of shift in $T^{MI}$ [$T^{MI}$(LBMO1400) = 335 K; $T^{MI}$(LBMO$Ag_{0.4}$) = 265 K] with Ag addition. On comparing peak resistivity ($\rho_{peak}$) in LBMO1300Ag & LBMO1400Ag series, we found decreased $\rho_{peak}$ values for LBMO1400Ag series (as can be seen from Fig.2(a) & 2(b)), which can be explained on the basis that high temperature sintering process enhances the grain growth [15,16]. That is to say that the sample prepared at higher temperature has better grain growth as well as better connectivity. It is worth noting that the XRD of LBMO1300Ag shows (Fig.1(a)) small quantity of BaO. On the other hand the LBMO1400Ag is nearly a phase pure series (Fig.1(b)). One can also describe the nature of conduction process in terms of the sharpness in metal-insulator transition, which has improved for silver added samples in present case. The sharpness of metal-insulator transition is described in terms of temperature coefficient of resistivity (TCR), defined as, $1/\rho \times (d\rho/dT) \times 100$,which is an important parameter in defining its application in infrared detectors. The peak TCR values are seen across the $T^{MI}$, i.e. across the metal insulator transition temperature. The plots of TCR for LBMO1300Ag and LBMO1400Ag are shown respectively in the insets of Fig. (2a) and 2(b). The peak TCR value for $Ag_{0.3}$($1300^0$C) is 1.8 % as compared to only 0.42% for pure LBMO1300. For LBMO:$Ag_{0.4}$(1400°C) and its pristine sample, the values for peak TCR are 2.84% and 1.5% respectively. The sharpness in transition or otherwise better TCR values can be explained on the basis that silver with low melting point has acted as a catalyst for improving the grain connectivity and grain growth [7,12,17,18], hence increasing the conduction process between the grains.

Fig. 2(c) depicts the variation of resistivity with temperature at zero field and at a field of 7 Tesla. The main panel is for pure LBMO(1400°C) and inset is for pure LBMO(1300°C) clearly demonstrating the decreases in resistivity under applied magnetic field for both pristine samples. As clearly seen maximum change $\Delta\rho$ ($\Delta\rho = \rho(0) - \rho(\mathrm{H})$ ,where $\rho(0)$ is resistivity at zero field and $\rho(\mathrm{H})$ is resistivity in applied field) is seen around transition temperature. Further, $\Delta\rho$ decreases with decrease in temperature and attains nearly constant value for temperature say below 30K .As far as LBMO:$Ag_x$ (x = 0.1,0.2,0.3,0.4) composites are concerned, maximum $\Delta\rho$ is again seen around transition temperature and the same decreases as temperature goes down, plots not shown. The shift of transition temperature ($T^{MI}$) with magnetic field as evident from fig. 2(c) is explained as follows: manganites near transition temperature are in mixed phase (paramagnetic-ferromagnetic) so application of external magnetic field suppresses paramagnetic phase and enhances the magnetic ordering i.e. alignment of spins which enhances conduction [3,4]. Another fact is that two peaks seen for zero field data of LBMO1300/1400 pristine samples are wiped out in their respective measurements under applied field. This is further explainable on the basis that the varying amounts of competing para-magnetic (PM) insulating and ferro-magnetic (FM) metallic components in the LBMO manganite gets more homogenized under magnetic field.

Fig. 3(a) & 3(b) describes the magneto-resistance (MR%) with temperature for LBMO1300Ag & LBMO1400Ag under applied field of 7 Tesla ($MR^{7T}$). Significant magneto resistance throughout the temperature range for all the samples is observed. Magneto-resistance (MR) is defined as, MR% = $((\rho(0) - \rho(\mathrm{H}))/\rho(0))$ x 100. We would like to discuss figure 3(a) & 3(b) in three separate regions. Region I consists of $MR^{7T}$ above $T^{MI}$. In this zone an increase in $MR^{7T}$ has been observed, which is due to suppression of paramagnetic phase prior to $T^{MI}$. Since the onset of ferromagnetic phase near $T^{MI}$ is accompanied with double exchange hence suppression of paramagnetic phase increases the number of conduction electron and thus resistivity is decreased with magnetic field [3,4].

Region II, is the region of about 10 K around the transition temperature ($T^{MI}$ ±5 K) where maximum magneto resistance ($MR^{7T}{}_{max}$) is achieved. $MR^{7T}{}_{max}$ is about 68% for LBMO:$Ag_{0.4}$ (1400°C) and 55 % for LBMO:$Ag_{0.4}$(1300°C). LBMO(1400°C) pure has $MR^{7T}{}_{max}$ of 41 % and that of LBMO(1300°C) is 38.8 % .The origin of high values of MR about transition as observed in thin flims of polycrystalline manganites is mainly due intrinsic MR which originates from ordering of magnetic moment in ferromagnetic regime. The spin polarization (odering) opens new channels for conduction and hence high MR values are obtained. On comparing $MR^{7T}{}_{max}$ of LBMO (1400°C) with that of its silver composites, a 24 % increment in $MR^{7T}{}_{max}$ has been observed with silver doping of x = 0.4 and 17% for Ag0.3 samples. Improved values of $MR^{7T}{}_{max}$ have been observed with silver for both LBMO1300Ag and LBMO1400Ag series. Improved MR in region II for silver composite can be explained on the basis that with silver the grain size (magnetic domain) increases, see the SEM results in Figs. 1(d) and (e). Increased magnetic domain size warrants larger spin polarization and hence higher MR at same fields. $MR^{7T}{}_{max}$ has also improved with increase in sintering temperature i.e. $MR^{7T}{}_{max}$ is 68% for LBMO:$Ag_{0.4}$(1400°C ) and 54.2% for LBMO:Ag0.4(1300°C). This further strengthens the point that the increased grains size/connectivity warrants the larger magnetic domains and hence larger MR.

Region III is the region below transition temperature .In this region we find that $MR^{7T}$ increases down up to 5 K. Let us first discuss pristine samples of LBMO1300Ag and LBMO1400Ag series in region III, as evident from fig. 3(a) & 3(b) $MR^{7T}$ of pristine samples at 5 K is higher than $MR^{7T}_{max}$ of region II. More precisely $MR^{7T}_{5K}$ is 41.6% (LBMO1300°C) and 42.7%(LBMO1400°C). Increase in low temperature high field $MR^{7T}$ basically in polycrystalline compounds is attributed to extrinsic MR, where grain boundary (GB) effect contributes substantially [19]. So high $MR^{7T}_{5K}$ in our pristine samples is mainly due to effective GB contribution. In the case of silver doped samples, $MR^{7T}_{5K}$ are nearly same for all samples and much less than $MR^{7T}_{max}$. As seen from our SEM results and discussed earlier, the grains morphology of LBMO:$Ag_x$ composites is much improved in terms of their size and connectivity in comparison to their pristine counter parts. This clearly shows that GB scattering effects are much less in LBMO:Ag samples than their pristine counterparts, and hence lower MR at 5 K. On the other hand the MR across $T^{MI}$ is more pronounced in LBMO:Ag composites simply due to better coupled FM domains and hence larger spin polarization effects under magnetic field, resulting in higher intrinsic MR.

The isothermal MR curves i.e. MR% vs field (± 7 Tesla) plots are shown in insets of Fig. (a) & 4(b) at various temperatures of 5, 100, 200, 300 and 400K. Insets, in Fig. 4(a) depict the isothermal plots for pure LBMO1300/1400 samples and the ones in 4(b) are for LBMO:$Ag_{0.4}$ (1400°C /1300°C). The MR% is nearly negligible and is U-type in shape at 400 K for both pure LBMO1300 and LBMO140, insets Fig. 4(a). This is expected because both the systems at these temperatures are in paramagnetic (PM) state. At 300, 200, 100 and 5K, the shapes of the isothermal MR plots is of V-type, with slight change in the shape for 5 K, where initially (low field) slope is steeper, which is in agreement with the fact that at such a lower temperature spin are more ordered and hence spin dependent scattering between the grains is reduced [19]. Interestingly the MR is maximum at 5 K for both pure LBMO1300 and LBMO1400, see Figs inset (4a). On the other hand for LBMO$Ag_{0.4}$ (1300) and 1400, the 5 K MR is much less (35%) than at 300 K (60%) in 7 Tesla applied field. This is basically due to the fact that the $T^{MI}$ of both LBMO$Ag_{0.4}$ (1300) and 1400 C samples is close to 300 K, and hence maximum MR at this temperature. It is clear from the isothermal MR results that the LBMO$Ag_{0.4}$ samples do have maximum MR at 300 K (insets Fig. 4b) and is nearly double to that as for pure LBMO at same temperature (insets Fig. 4a).

Figure 5 depicts the magnetization of LBMO1400$Ag_x$ samples in field-cooled situation under applied field of 1000 Oe in the temperature range of 400 down to 5K. As shown in figure 5, magnetization decreases with silver percentage; further saturation moment at 5 K is above 80 emu/g for LBMO (1400°C) and it decreases, though not linearly, with silver. Inset of fig. 5, are curves of derivative of magnetization as a function of temperature with their minima shifting towards lower values. This tells us that the paramagnetic-ferromagnetic transition temperature ($T_C$) decreases with increase in Ag content. This is in agreement with the shift in $T^{MI}$ (Figs. 2a &b), mentioned in resistivity measurements section. This also explain the M(H) behavior where LBMO:$Ag_{0.4}$ is paramagnetic at 300 K, since $T_C$ is below 300 K (Fig.

4b). In case of LBMO1400pure, since the $T_C$ is above 300 K, the M(H) plot clearly exhibit the fasrromagnetism, see Fig. 4(a). This shows that the $T^{MI}$ (resistiviy measurements, Figs. 2 a & b) and $T_C$ (magnetization measurements, Figs. 4 a, b and 5) are in agreement with each other, i.e., PM to FM transition temperature decreases with increase in Ag content.

## Summary and Conclusion

Summarily, we synthesized $La_{0.7}Ba_{0.3}MnO_3$(LBMO):$Ag_x$ ( x = 0, 0.1, 0.2, 0.3, 0.4 ) composites with sintering temperature 1300°C (LBMO1300Ag)and 1400°C(LBMO1400Ag) .Improved magneto-resistance($MR^{7T}$) and TCR values are obtained for silver added samples. We are reporting MR as high as 68% for LBMO:$Ag_{0.4}$(1400°C) at 300 K. On comparing LBMO1300Ag with LBMO1400Ag we found that samples sintered at high temperature are of slightly better quality. Interestingly we also found that the addition of silver in LBMO has bought changes in transition temperature, which does not occur in the case of LCMO:$Ag_x$ composites. High value (68%) of MR at room temperature (300K) for LBMO:Ag composites may find its place in practical applications.

## Acknowledgement

Authors would like to thank Mr. K. N. Sood from NPL for the SEM micrographs. Authors from NPL appreciate the interest and advice of Prof. Vikram Kumar (Director) NPL in the present work. One of us Rahul Tripathi acknowledges the financial help in form of JRF-NET fellowship from the UGC-India.

## Figure Captions

Fig.1(a). X-ray Diffraction pattern of $La_{0.7}Ba_{0.3}MnO_3$:$Ag_x$ (1300°C) (x = 0.0,0.1,0.2,0.3,0.4) at room temperature, inset shows enlarge region between 2(θ) = 57-59$^0$

Fig.1(b). X-ray Diffraction pattern of $La_{0.7}Ba_{0.3}MnO_3$:$Ag_x$ (1400°C) (x = 0.0,0.1,0.3,0.4) at room temperature, inset show X-ray enlarged diffraction pattern of $La_{0.7}Ba_{0.3}MnO_3$ (1300°C) and $La_{0.7}Ba_{0.3}MnO_3$ (1400°C) at room temperature between 2(θ) = 66-69$^0$

Fig.1(c). Figure shows variation of lattice parameter a, b and c with silver content for LBMO:$Ag_x$1400 $^0$C.

Fig.1(d). SEM micrograph of $La_{0.7}Ba_{0.3}MnO_3$ (1400°C) polycrystalline sample

Fig.1(e). SEM micrograph of $La_{0.7}Ba_{0.3}MnO_3$:$Ag_{0.4}$ (1400°C) polycrystalline sample

Fig.2(a). Normalized resistivity as a function of temperature for $La_{0.7}Ba_{0.3}MnO_3$:$Ag_x$ (1300°C), Inset shows TCR % Vs Temperature Plots of $La_{0.7}Ba_{0.3}MnO_3$:$Ag_x$ (1300°C)

Fig.2(b). Normalized resistivity as a function of temperature for $La_{0.7}Ba_{0.3}MnO_3$:$Ag_x$ (1400°C), Inset shows TCR % Vs Temperature Plots of $La_{0.7}Ba_{0.3}MnO_3$:$Ag_x$ (1400°C)

Fig.2(c). Resistivity as a function of temperature with field (7 Tesla) and without field for $La_{0.7}Ba_{0.3}MnO_3$ (1400°C), inset shows resistivity as a function of temperature with field ( 7 Tesla ) and without field for $La_{0.7}Ba_{0.3}MnO_3$ (1300°C)

Fig.3(a). Figure shows magneto resistance (MR) as a function of Temperature for $La_{0.7}Ba_{0.3}MnO_3$:$Ag_x$ (1300°C)

Fig.3(b). Figure shows magneto resistance (MR) as a function of Temperature for $La_{0.7}Ba_{0.3}MnO_3$:$Ag_x$ (1400°C)

Fig.4(a). Figure shows M-H curves of $La_{0.7}Ba_{0.3}MnO_3$ (1400°C) Inset: MR as a function of applied field(± 7 T) of $La_{0.7}Ba_{0.3}MnO_3$ (1400°C) & $La_{0.7}Ba_{0.3}MnO_3$ (1300°C)

Fig.4(b). Figure shows M-H curves of $La_{0.7}Ba_{0.3}MnO_3$:$Ag_{0.4}$ (1400°C) Inset : MR as a function of applied field(± 7 T) of $La_{0.7}Ba_{0.3}MnO_3$:$Ag_{0.4}$ (1400°C) & $La_{0.7}Ba_{0.3}MnO_3$:$Ag_{0.4}$ (1300°C)

Fig.5. Figure shows magnetization as a function of temperature of $La_{0.7}Ba_{0.3}MnO_3$:$Ag_x$ (1400°C),inset shows the variation of derivative dM/dT with temperature for $La_{0.7}Ba_{0.3}MnO_3$:$Ag_x$ (1400°C)

Figure 1(a).

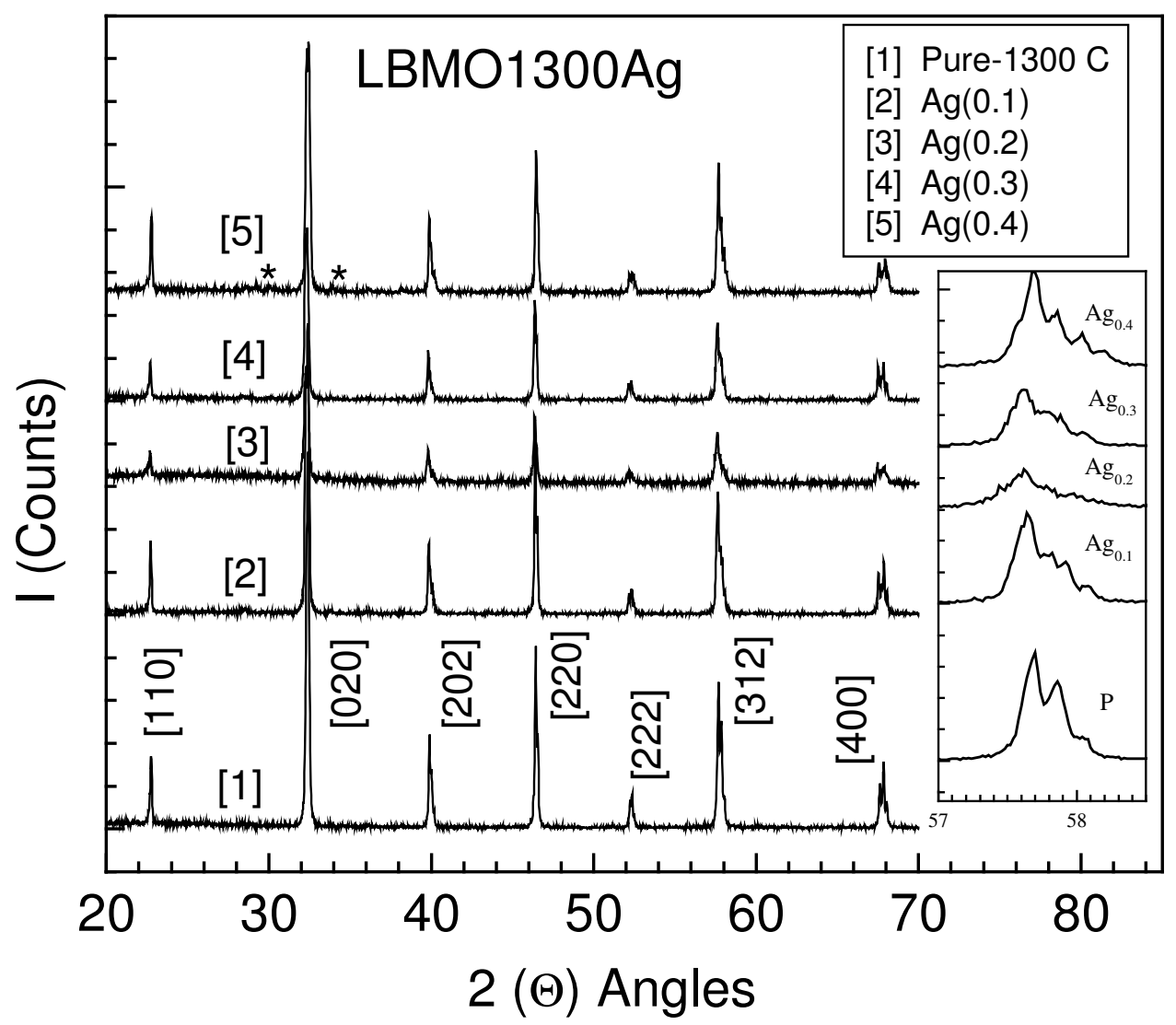

Figure 1(b)

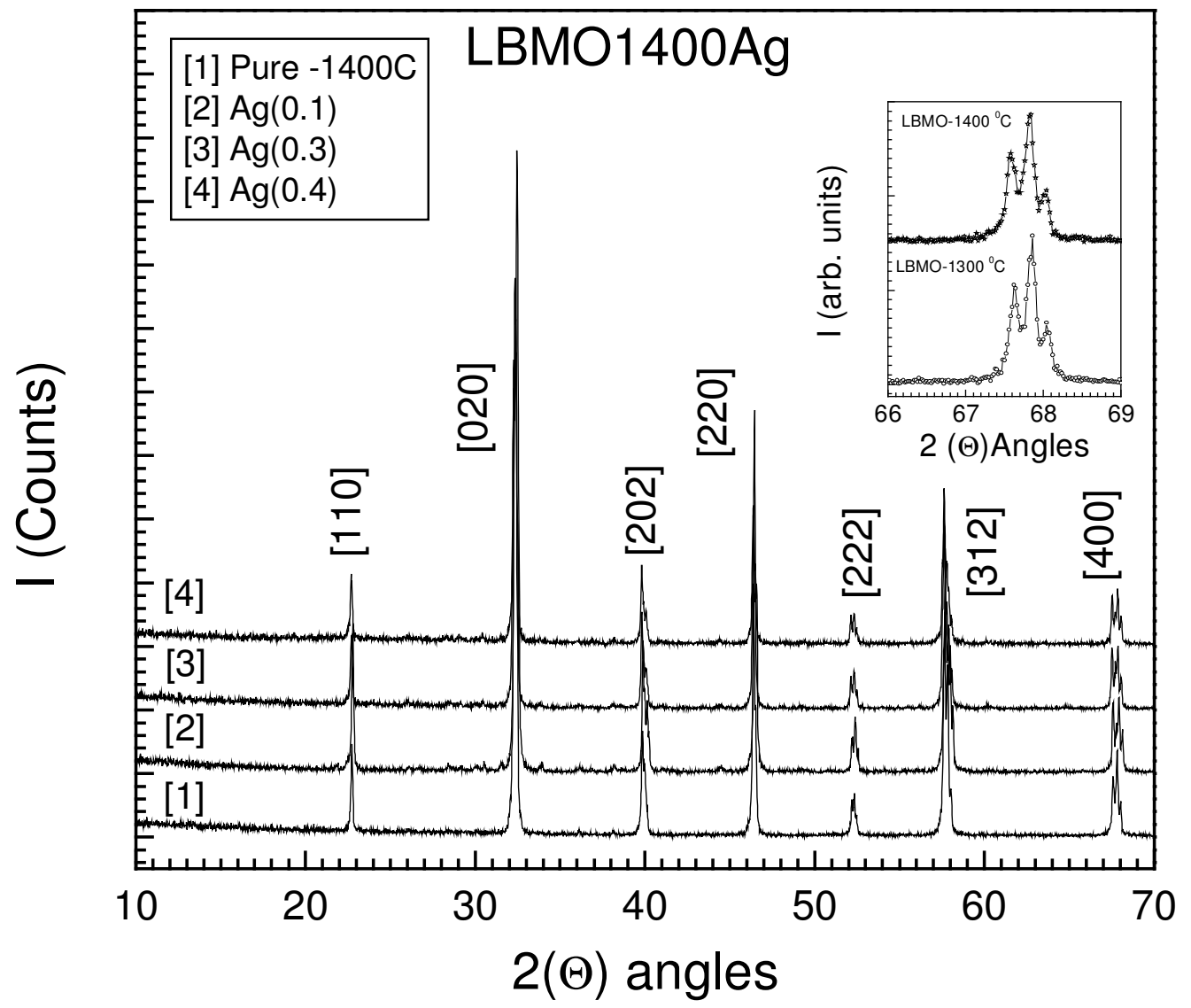

Figure 1(c)

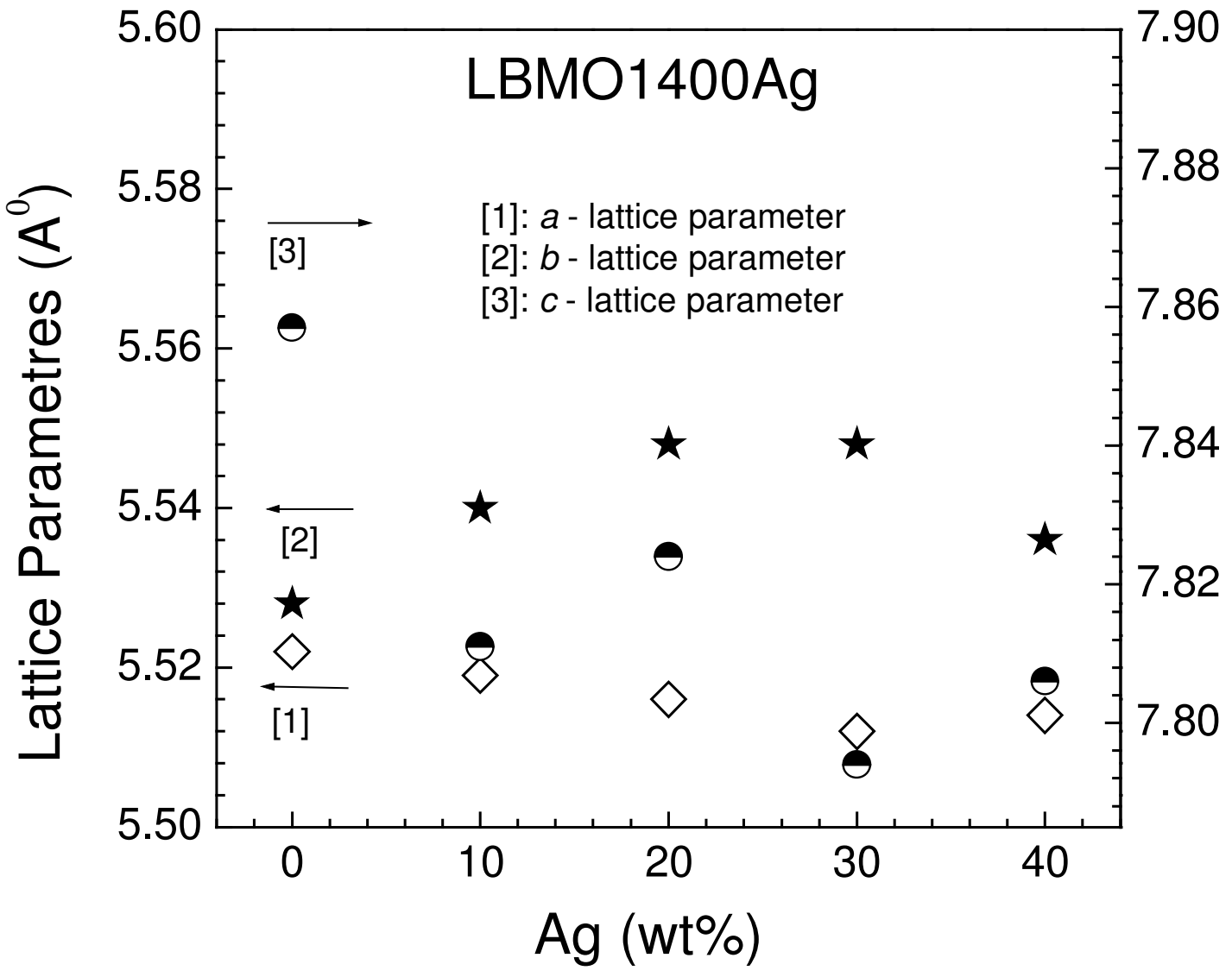

Figure 1(d)

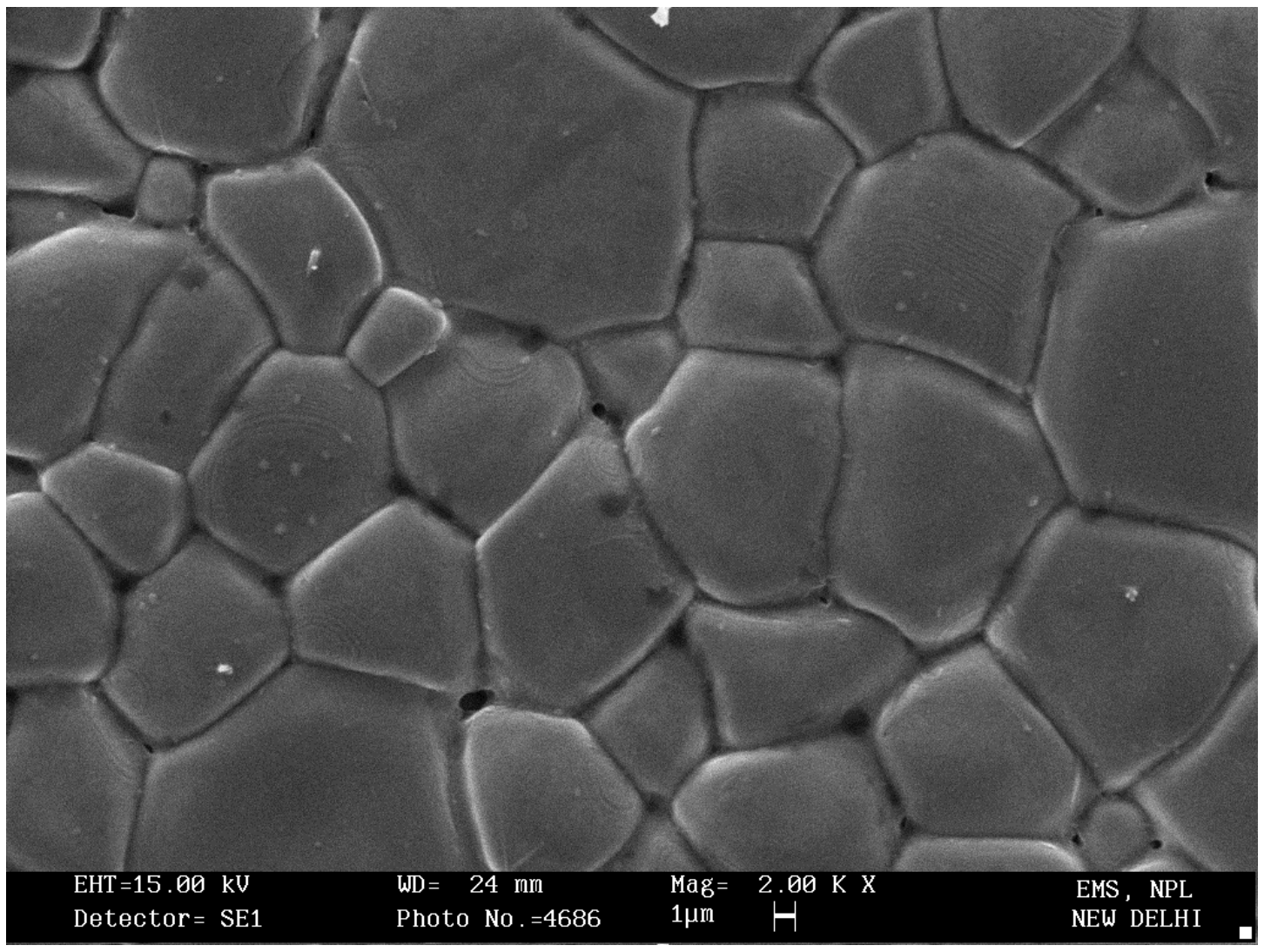

Figure 1(e)

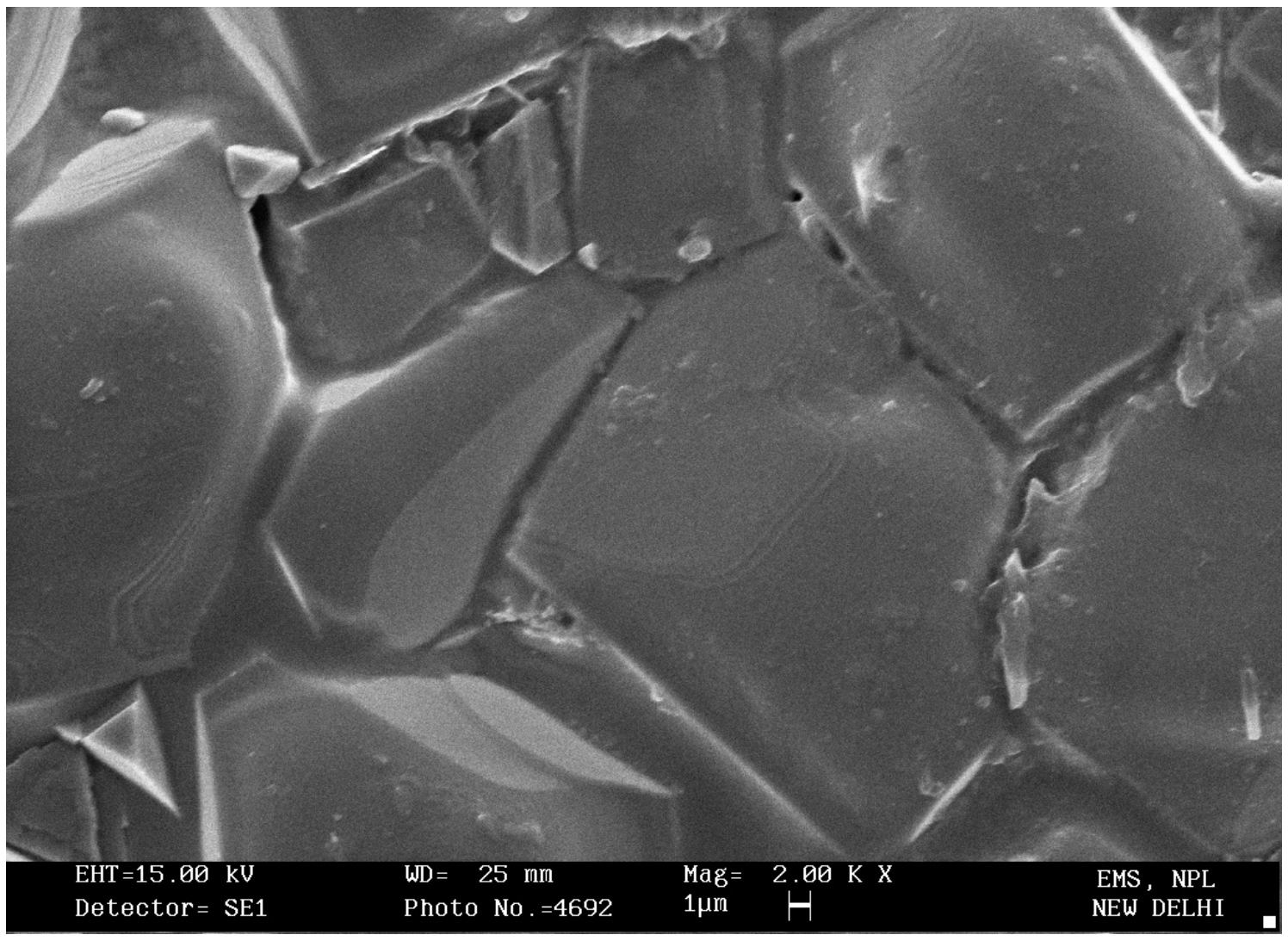

Figure 2(a)

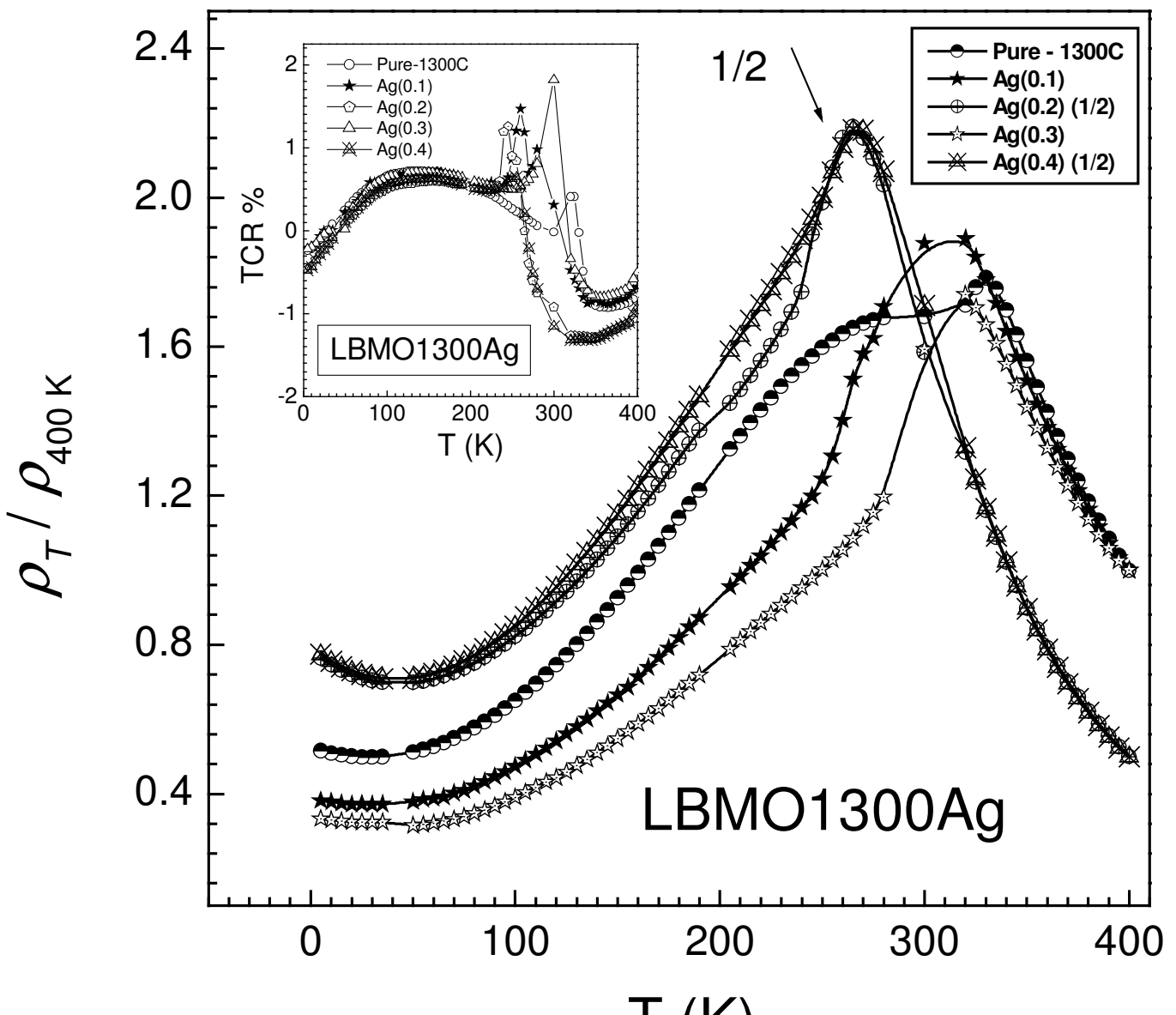

Figure 2(b)

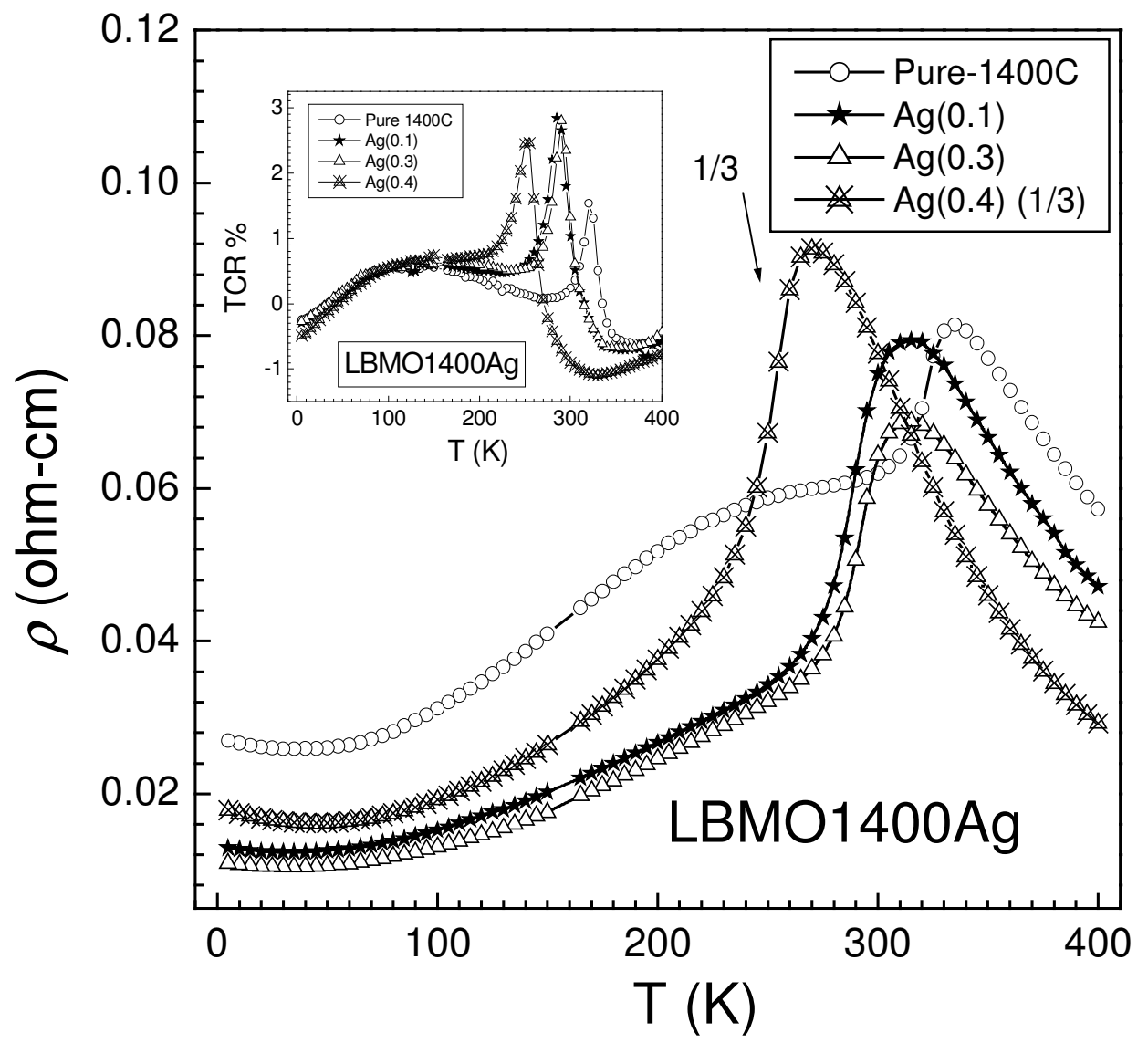

Figure 2(c)

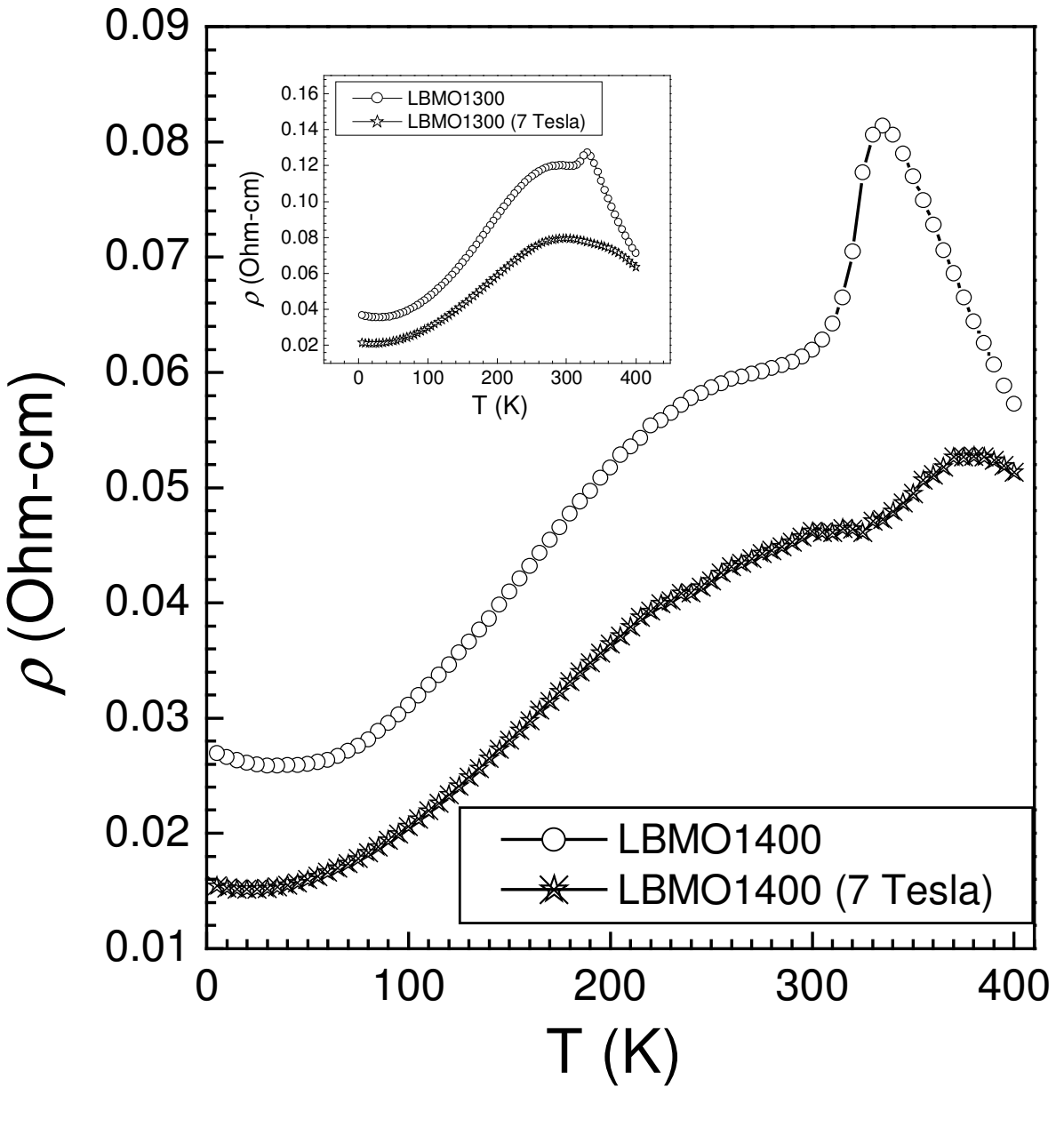

Figure 3(a)

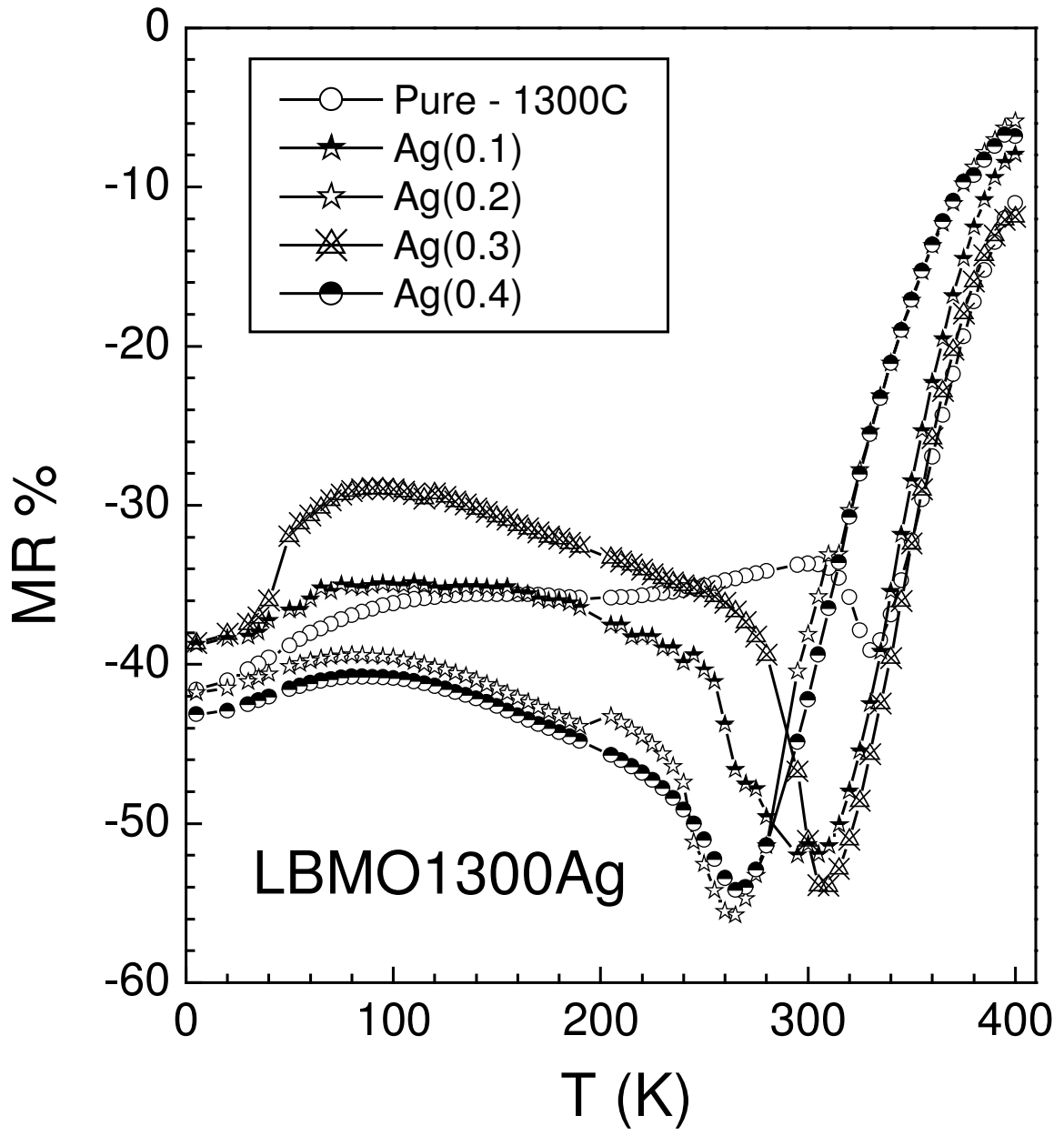

Figure 3(b)

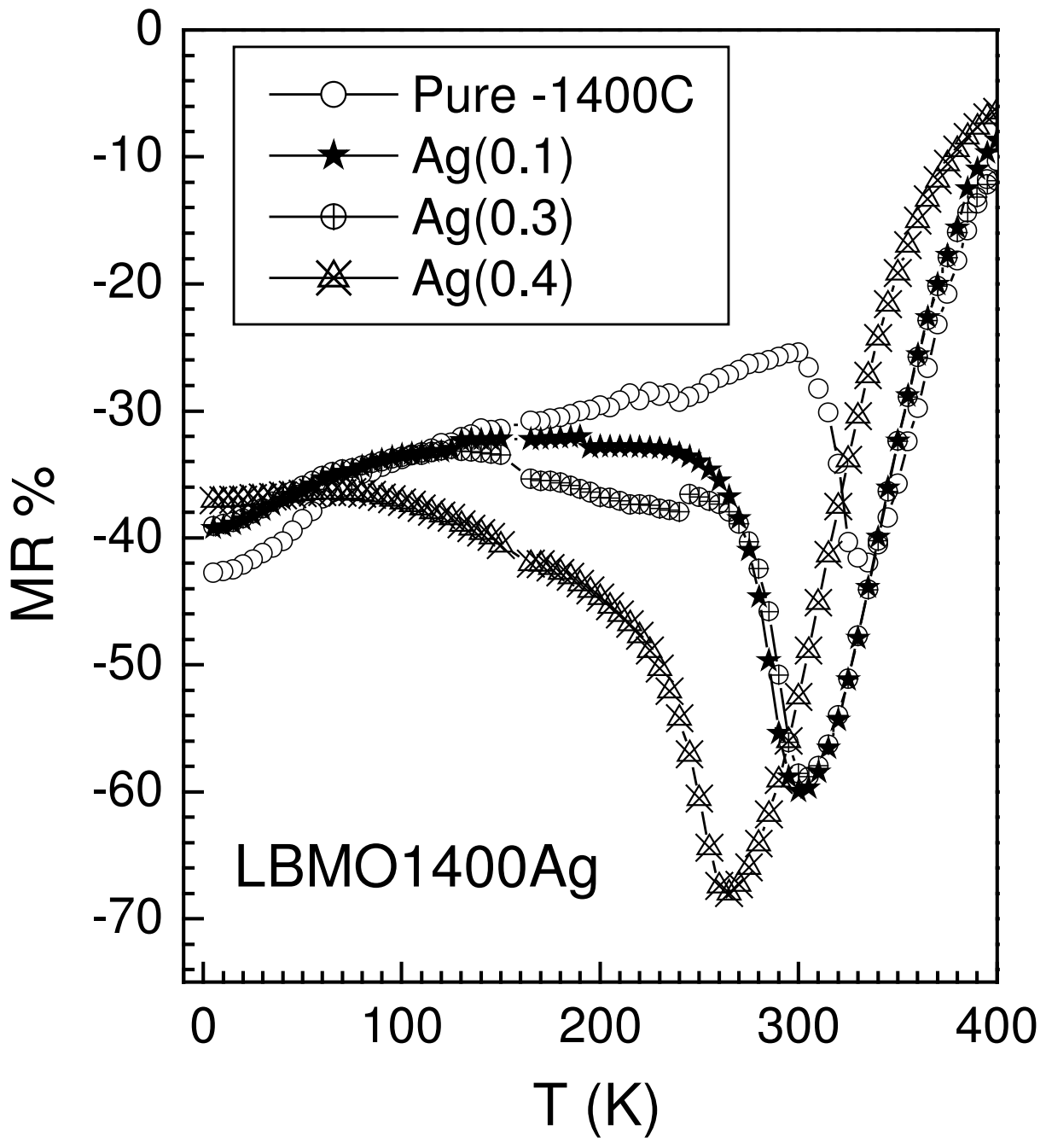

Figure 4(a)

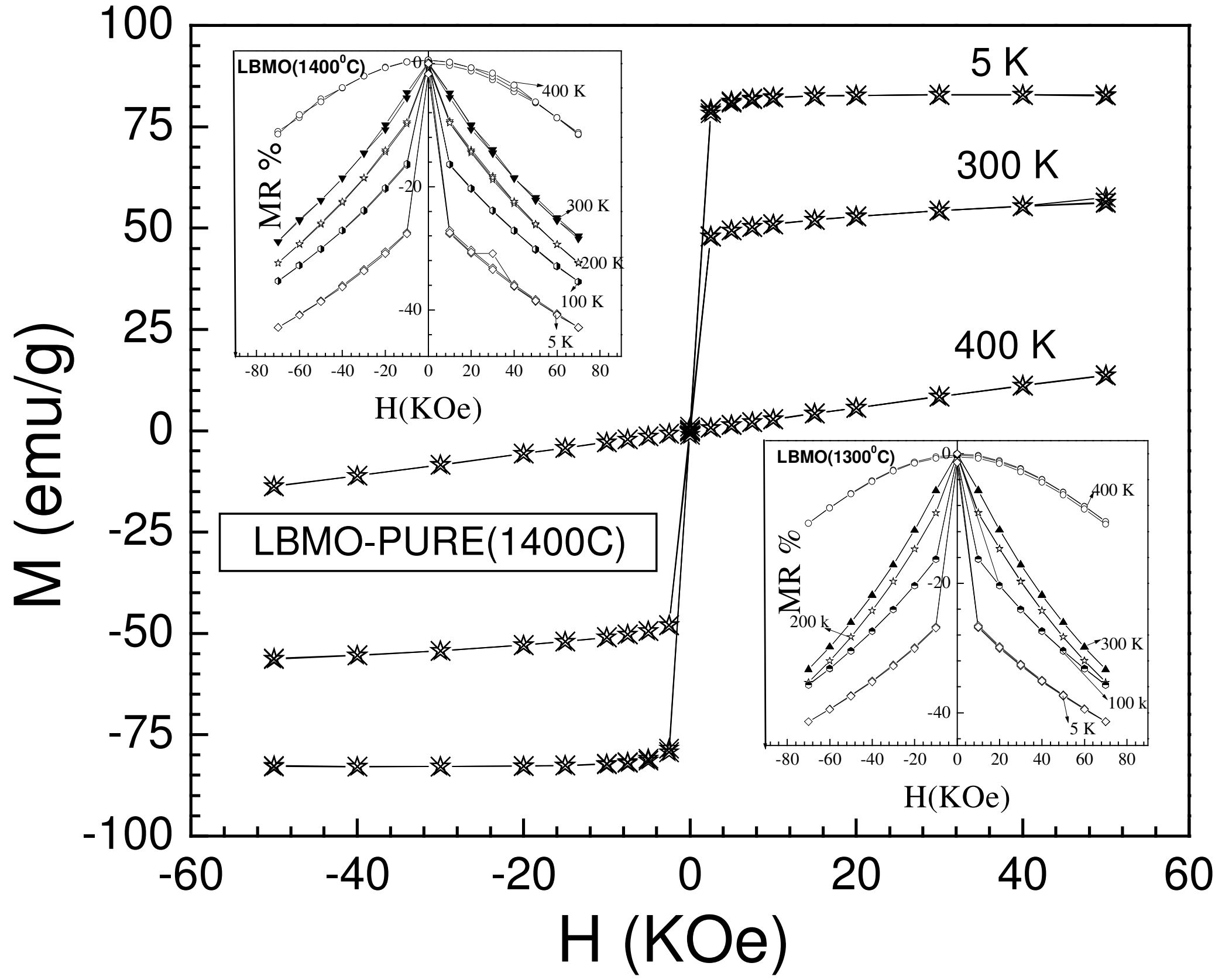

Figure 4(b)

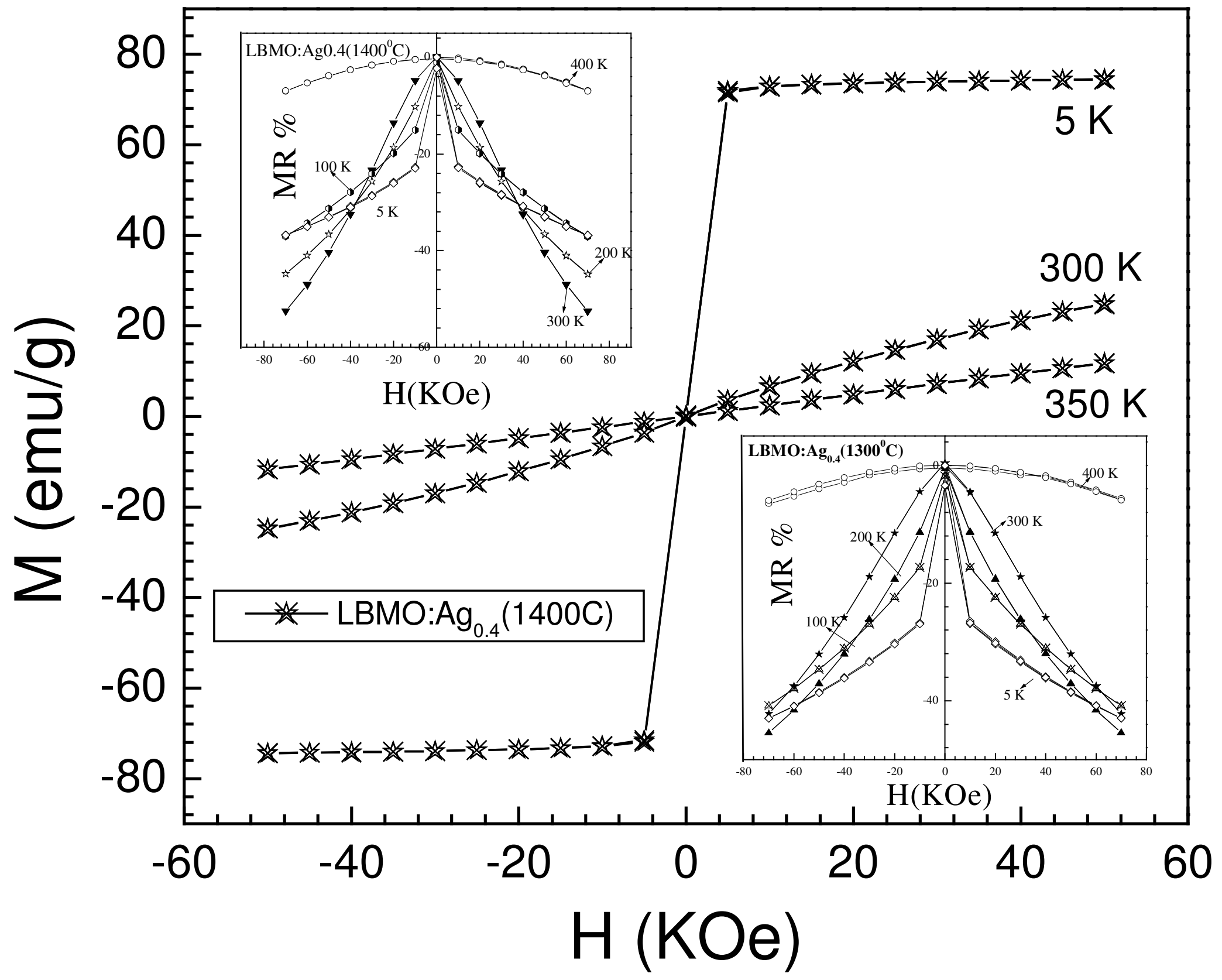

Figure 5

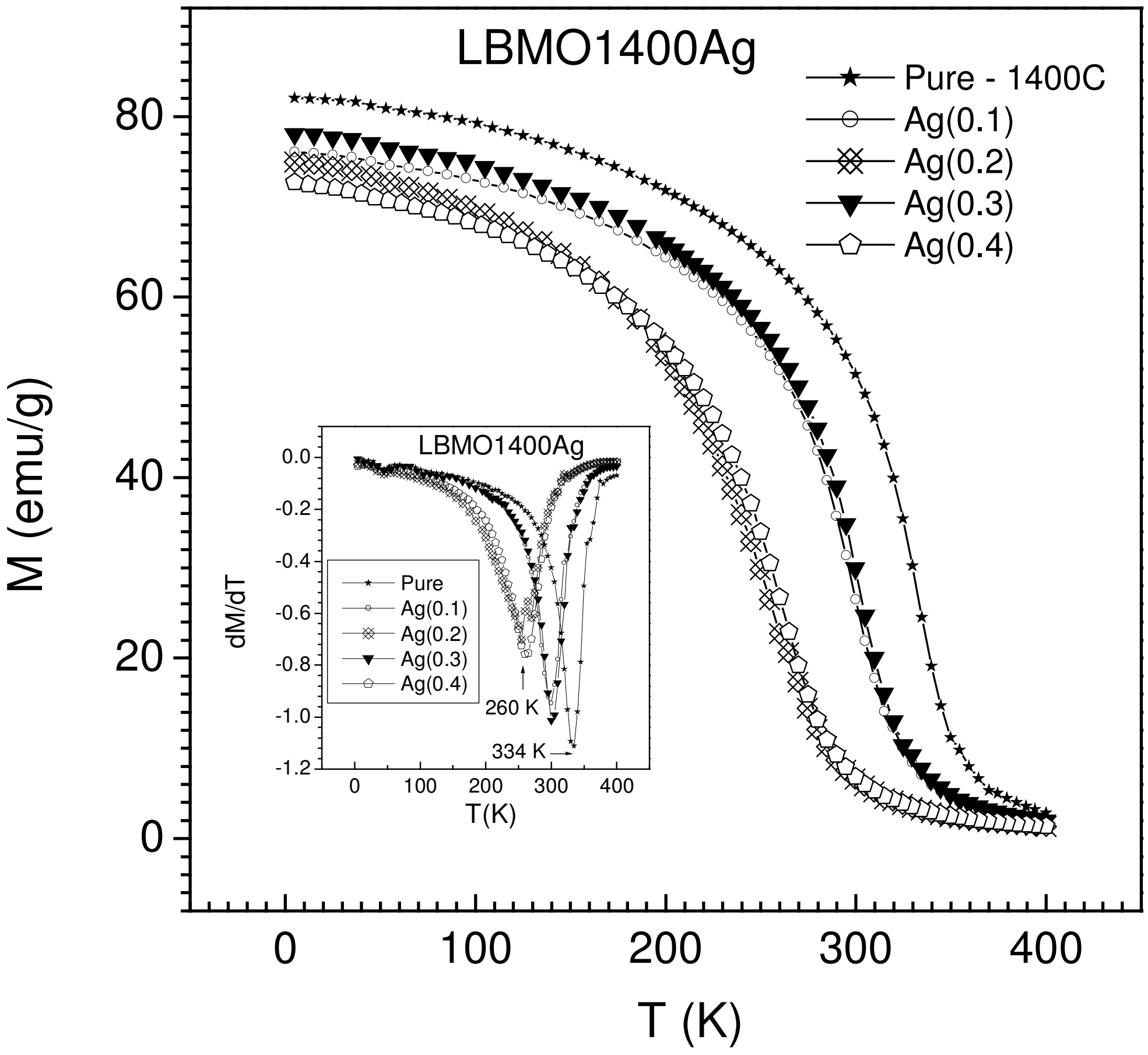